\documentclass[showpacs,preprintnumbers,twocolumn,amsmath,amssymb]{revtex4}
\usepackage{graphicx}% Include figure files
\usepackage{dcolumn}% Align table columns on decimal point
\usepackage{bm}% bold math

\begin{document}
%
%\newpage

%\preprint{APS/123-QED}

\title{An EPR local theory with local correlation}
\author{W. LiMing} \email{wliming@scnu.edu.cn}
\author{Zhilie Tang} \affiliation{Dept. of
Physics, and Institute for Condensed Matter Physics, School of
Physics and Telecommunication Engineering, South China Normal
University, Guangzhou 510006, China} 
\date{\today}
\begin{abstract}
An EPR local theory with local correlation is proposed to give an
explanation for the contractions in the GHZ-like schemes for Bell's
theorem and the violation to Bell's inequality. It agrees with the
experimental predictions for the GHZ state of three entangled spins,
the entangled state of two spins and one spin state. The
contradiction in the GHZ-like schemes and the violation to Bell's
inequality can be attributed to the local correlation between the
EPR elements of physical reality.
\end{abstract} 
 \maketitle

\section{Introduction}
Einstein, Podolsky, and Rosen (EPR) proposed a deterministic local
theory\cite{Einstein}, elements of physical reality, to reproduce
the predictions of quantum theory. This local theory was thought to
be refuted by Bell's theorem, which demonstrated that the
interpretations of quantum theory must be nonlocal by the well known
Bell's inequality\cite{Bell}. Greenberger, Horne, and Zeilinger
(GHZ)\cite{GHZ} provided an experimental scheme to test Bell's
theorem without using inequalities. Because this scheme requires at
least three observers for an entangled state of three particles with
spin-$1/2$, Hardy gave a proof with an entangled state of only two
spins\cite{Hardy}, but valid only for non-maximally entangled
states. Then Cabello provided a proof for two observers using two
pairs of maximally entangled particles based on Hardy's
criterion\cite{Cabello2}  and another one on GHZ's
criterion\cite{Cabello}. Following Cabello's latter idea Chen {\it
et al} proved Bell's theorem by one pair of two entangled particles
with two spin degrees of freedom and two space degrees of
freedom\cite{Chen}.

 The series of proofs of
GHZ\cite{GHZ}, Hardy\cite{Hardy}, Cabello\cite{Cabello},
Chen\cite{Chen} and the relevant experiments have similar criterion,
thus are called the GHZ-like schemes latter in this paper. They
found a contradiction between EPR's elements of physical reality and
the predictions of experiments, hence claimed that elements of
physical reality do not exist thus local theories do not hold. The
contradiction occurs when $\sigma_x$ and $\sigma_y$ are assigned
\textit{real values}, $m_x, m_y$, though which are called
\textit{elements of physical reality}. In quantum theory, however,
one has strong evidences that it is impossible to assign values to
$\sigma_x$ and $\sigma_y$ since they do not commute each other. This
implies that a local theory may survive if it works together with a
local correlation between $m_x$ and $m_y$.

In this paper we first analyze the contradiction revealed by the GHZ
scheme. It is found that this contradiction can be explained by the
local correlation between EPR's elements of physical reality. We
then demonstrate that the local correlation satisfies all the
predictions of GHZ,two-spin entangled state and even one spin state.
It is shown that the all GHZ-like schemes can be explained by the
local correlation. Finally we point out that Bell's inequality is
violated only in the region of the local correlation and thus is
attributed to the local correlation. A conclusion is given in the
last section that an EPR local theory with the local correlation
survives.

\section{GHZ's proof is trivial}\label{predictions}
GHZ\cite{GHZ} considered an entangled state of three particles, A, B
and C with spin-${1\over 2}$
\begin{align}
|\Psi\rangle &= {1 \over \sqrt
2}(|\uparrow_A\uparrow_B\uparrow_C\rangle -
|\downarrow_A\downarrow_B\downarrow_C\rangle),
\end{align}
where $|\uparrow\rangle$ and $|\downarrow\rangle$ stand for spin up
and down states respectively. It is easy to check
\begin{align}\label{xyz1}
\sigma_x^A\sigma_y^B\sigma_y^C|\Psi\rangle &=
|\Psi\rangle,\\\label{xyz0}
\sigma_y^A\sigma_x^B\sigma_y^C|\Psi\rangle &= |\Psi\rangle,\\
\sigma_y^A\sigma_y^B\sigma_x^C|\Psi\rangle &= |\Psi\rangle,\\
\sigma_x^A\sigma_x^B\sigma_x^C|\Psi\rangle &= -|\Psi\rangle,
\label{xyz4}
\end{align}
where $\sigma_{x,y}$ are  Pauli matrices.  The GHZ state is the
common eigenstate of the four commuting Hermitian operators
$\sigma_x^A\sigma_y^B\sigma_y^C,
\sigma_y^A\sigma_x^B\sigma_y^C,\sigma_y^A\sigma_y^B\sigma_x^C$, and
$\sigma_x^A\sigma_x^B\sigma_x^C$ with eigenvalues $1,1,1,-1$,
respectively. Since these operators commute with each other, they
can be observed simultaneously. These quantities can only be locally
observed, e.g., $\sigma_x^A\sigma_y^B\sigma_y^C$ is observed by
measuring $\sigma_x^A$, $\sigma_y^B$ and $\sigma_y^C$ independently.
Since the product of the three measured values is certainly equal to
the eigenvalue of $\sigma_x^A\sigma_y^B\sigma_y^C$, two of them
predict the other. EPR's criterion tried to explain this nonlocal
prediction by means of a local theory, elements of physical
reality\cite{Einstein,GHZ,Mermin}. In the present case the elements
satisfy
\begin{align}\label{mmm1}
m_x^A\,\,m_y^B\,\,m_y^C &= 1,\\ \label{mmm2}
m_y^A\,\,m_x^B\,\,m_y^C&= 1,\\\label{mmm3} m_y^A\,\,m_y^B\,\,m_x^C&= 1,\\
\label{mmm4}m_x^A\,\,m_x^B\,\,m_x^C&= -1,
\end{align}
where $m_{x,y}^L = \pm 1, L= A,B,C$. One obtains a contradiction
$1=-1$ when multiplying the above four equations side by side. Due
to this contradiction GHZ claimed that Bell's theorem without
inequalities was proved, and no element of physical reality might
exist.

The precondition of this proof is the independent existence of the
elements of physical reality, which are determined by a set of
hidden variables $\lambda =(\lambda_1,\lambda_2,....)$
\begin{align}\label{hypo}
m_x = m_x(\lambda), \,m_y = m_y(\lambda).
\end{align}
In quantum theory, however, it is well known  that since $\sigma_x$
and $ \sigma_y$ do not commute each other, (\ref{hypo}) certainly
gives rise to problems. For example, due to the following equality
\begin{align}\label{QT}\sigma_x\sigma_y=-\sigma_y\sigma_x,
\end{align}
(\ref{hypo}) gives an ill equation $m_x m_y=-m_y m_x$ if the Pauli
matrices are valuated  $m_x$ and $m_y$, respectively. In the point
of view of experiments, it is also well known that $\sigma_x$ and $
\sigma_y$ cannot be determined simultaneously. A schematic scheme
for testing this conclusion is shown in FIG.1.
\begin{figure}
\includegraphics[width=7cm,height=5cm]{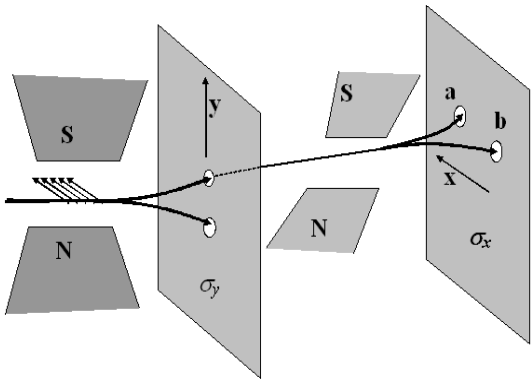}
\caption{Schematic diagram for measuring electron's spin components
in the x- and y-directions by the Stern-Gerlach method. The spins of
the incident polarized electrons point to the x-axis direction. The
magnetic field in the left-hand side is in the vertical(y-)
direction, and that in the right-hand side is in the x-direction.}
\end{figure}

This scheme is to measure $\sigma_x$ and $\sigma_y$ of electrons
using the Stern-Gerlach method. The incident electrons are polarized
in the x-axis direction as shown in the figure. They are supposed to
have an element of physical reality $m_x =+1$ since a measurement to
$\sigma_x$ gives a certain value $+1$. When these polarized
electrons pass through the first nonuniform magnetic field in the
y-axis direction they split into two branches. Choose the upper
branch, which determines the value of $\sigma_y$, i.e., $m_y = +1$.
The values $m_x$ and $m_y$ seem to be both determined at this stage.
When this branch passes the following nonuniform magnetic field,
however, it again splits into two branches, arriving at windows $a$
and $b$ equally, i.e., the value of $m_x$ is missing. It is seen
that in fact the elements of physical reality, $m_x$ and $m_y$, do
not exist simultaneously.
 Therefore, the hypothesis (\ref{hypo}) does not
hold. The goal of GHZ's scheme is achieved by means of such a simple
test.

Hence either in the aspect of quantum theory or experiment GHZ's
conclusion is quite trivial.
%This contradiction may have another fashion. Noting that the three
%particles in the GHZ state are on an equal footing, thus
%$m_x^A\,\,m_y^A = m_x^B\,\,m_y^B = m_x^C\,\,m_y^C$, then multiplying
%(\ref{mmm1}-\ref{mmm4}) gives
%\begin{align}\label{mm1}
%(m_x^A\,\,m_y^A)^6 =(m_x^B\,\,m_y^B)^6 =(m_x^C\,\,m_y^C)^6 =-1
%\end{align}
%This equation also contradicts the values $m_{x,y} = \pm 1$.
%
% This contradiction even exists on
%one particle. Consider the following Hermitian operator
%\begin{align}\label{ss}
%-i\sigma_x \,\sigma_y  = \begin{pmatrix} 1 & 0 \\ 0 & -1
%\end{pmatrix}
%\end{align}
%It is an observable quantity with eigenvalues $\pm 1$. If GHZ's
%elements of physical reality exist one has
%\begin{align}\label{mm3}
%-im_x\, m_y =\pm 1
%\end{align}
%This equation contradicts the values $m_{x,y}=\pm 1$.
%
%It is seen that GHZ's contradiction exists not only on entangled
%states of two or more particles but also on one particle.

\section{An EPR Local theory with local correlation}

The co-existence hypothesis (\ref{hypo}), however, is not a
necessity of a local theory. A local theory denies only nonlocal
correlations between different particles apart but it may admit a
local correlation between the elements of physical reality of one
particle. For example, in the case of the GHZ scheme the following
local correlation for each particle can be accepted by a local
theory,
\begin{align}\label{hypo1}
  m_{xy} \equiv m_x m_y = \pm i.
\end{align}
It assumes that when  $m_x$ is determined, {\it e.g.} $m_x =1$, then
$m_y$  will be stochastic or illy defined, and {\it vise versa}. In
fact, (\ref{QT}) has told that there must be a local correlation
between $m_x$ and $m_y$ if they did exist.
%This local correlation is
%in accordance with the prediction in the above experimental testing
%scheme.

Surprisingly, the local correlation (\ref{hypo1}) satisfies the
prediction equations in the GHZ scheme. For example, multiplying
(\ref{mmm1})-(\ref{mmm4}) one obtains
\begin{align}\label{mm1}
(m_x^A\,\,m_y^A)^2 (m_x^B\,\,m_y^B)^2 (m_x^C\,\,m_y^C)^2 =-1.
\end{align}
Substituting (\ref{hypo1}) into the above equation gives $-1 = -1$.
GHZ's contradiction disappears!
%
%It is also true in the two-particle entangled state and the single
%particle case. Multiplying (\ref{mxmx}) and (\ref{mxmx1}) one
%obtains
%\begin{align}
%(m_x^A\,m_y^A) (m_x^B\,m_y^B)&= -1
%\end{align}
%which also agrees with (\ref{hypo1}).
The local correlation gives an explanation to the contradiction in
GHZ's proof. It revives EPR's local theory.

One should not take the elements of physical reality, $m_x$ and
$m_y$, as really-measured values. From the point of view of
experiment one never tested Eqs.(\ref{xyz1}-\ref{xyz4}) on one group
of entangled particles.  When one of them, e.g., (\ref{xyz1}), is
realized through local measurements, i.e., $\sigma_x^A$,
$\sigma_y^B$ and $\sigma_y^C$ are measured independently, the GHZ
state $|\Psi\rangle$ collapses to the eigenstate of $\sigma_x^A$,
$\sigma_y^B$ and $\sigma_y^C$, e.g.,
$|\nwarrow\nearrow\nearrow\rangle$, here
$|\nwarrow\rangle,|\nearrow\rangle$ denote the eigen-states of
$\sigma_x$, $\sigma_y$. Thus one has to pick new groups of entangled
particles in the GHZ state to realize other equations. Different
groups, however, give different measured values. Therefore, it is
impossible to represent different measured values on different
groups of particles by the same group of numbers $(m_x,\, m_y)$.
This is why $(m_x,\, m_y)$ are named only as elements of physical
reality.

%The elements, $m_x$ and $m_y$, originate in a free selection from
%the local measurements for Eqs.(\ref{xyz1}-\ref{xyz4}).  One should
%bear in mind, however, that $\sigma_x$ and $ \sigma_y$  do not
%commute each other. Once the value of $\sigma_x$ has been measured,
%the value of $\sigma_y$ will be stochastic. This claim has strong
%evidences in the history of quantum theory. Thus only the elements
%of physical reality in one of the equations,
%(\ref{mmm1}-\ref{mmm4}), can become really-measured values and
%others exist in our mind only.
Nevertheless, any form of elements of physical reality, even though
including the above local correlation, is definitely contradicting
to quantum theory. It is impossible to make two independent numbers,
either real or complex, to satisfy the anti-commuting relation,
(\ref{QT}). Therefore, a real experimental scheme to test local
theories should be able to refute the elements of physical reality
with the local correlation, such as (\ref{hypo1}), but not the
independent existence (\ref{hypo}). Unfortunately, neither Bell's
5inequality\cite{Bell}, GHZ's\cite{GHZ}, Hardy's\cite{Hardy,Gold},
5Cabello's\cite{Cabello}, and Chen's\cite{Chen} schemes, nor the
relevant experimental demonstrations did.
%They in fact refuted only
%the latter but not EPR with the local correlation.

\section{Local correlation in Hardy's proof}

In a general understanding to entanglement a maximally entangled
state should be most nonlocal. It is  surprising that Hardy's proof
is valid only for two \textit{non-maximally} entangled
particles\cite{Hardy, Gold}. Up to date nobody has provided a proof
for a maximally entangled state of two particles. Why is this the
case? The local correlation gives an answer.

Goldstein provided a simpler version\cite{Gold} For Hardy's proof.
He considered an entangled state of two particles,
\begin{align}|\psi\rangle
= a |\downarrow_1\downarrow_2\rangle + b
|\uparrow_1\downarrow_2\rangle + c|\downarrow_1\uparrow_2\rangle,
\end{align}
where $|\uparrow\rangle$ and $|\downarrow\rangle$ are two orthogonal
and normalized basis vectors. The four Hermitian operators, $\{\hat
U_i = |\uparrow_i\rangle \langle \uparrow_i|, \hat W_i =
|\beta_i\rangle \langle \beta_i|, i = 1,2\}$ are measured, where
\begin{align}|\beta_i\rangle
= {a |\downarrow_i\rangle + x |\uparrow_i\rangle \over \sqrt{|a|^2 +
|x|^2}},\quad x =\begin{cases} b \text{ for } i = 1\\ c
 \text{ for } i = 2\end{cases}.\end{align}
It was proved that the elements of physical reality, $ U_i, W_i$,
corresponding to $\hat U_i, \hat W_i$, contradict each other under
the condition $abc \neq 0$. This condition makes $|\psi\rangle$
non-maximally entangled.

We find that this contradiction is just the requirement of the local
correlation, because
\begin{align}
[\hat U_i, \hat W_i] = {a^* x |\uparrow_i\rangle \langle
\downarrow_i|-ax^*|\downarrow_i\rangle \langle \uparrow_i| \over
\sqrt{|a|^2 + |x|^2}}.
%\biggl(a^* x |\mu_i\rangle \langle \nu_i|-ax^*|\nu_i\rangle
%\langle \mu_i|\biggr)
\end{align}
When $a\neq 0, x\neq 0$ (i.e., $b\neq0, c \neq 0$) the operators
$\hat U_i$ and $\hat W_i$ do not commute each other, thus the local
correlation between $U_i$ and $W_i$ exists just under the condition
$abc \neq 0$. This is why  Hardy's proof does not fit the maximally
entangled state. In the point of view of experiment it is impossible
to measure $\hat U_i$ and $\hat W_i$ simultaneously when $abc \neq
0$. In this case $U_i$ and $W_i$ do not exist independently for a
local theory, i.e., there is a local correlation between them. Hence
Hardy's contradiction does not exist.

\section{maximally entangled state of two particles}
Although until now nobody refuted elements of physical reality by a
two-particle maximally entangled state, we find that the local
correlation fit this case. As an example, we consider a Bell-basis
state of two spin-${1\over 2}$ particles
\begin{align}
|\Phi^-\rangle &= {1 \over \sqrt 2}(|\uparrow_A\uparrow_B\rangle -
|\downarrow_A\downarrow_B\rangle).
\end{align}
One has
\begin{align}\label{xyz}
\sigma_x^A\sigma_x^B|\Phi^-\rangle &= -|\Phi^-\rangle,\\
\sigma_y^A\sigma_y^B|\Phi^-\rangle &= |\Phi^-\rangle,
\end{align}
where $\sigma_x^A\sigma_x^B$ and $\sigma_y^A\sigma_y^B$ commute each
other and can be experimentally determined simultaneously. In the
same way as the GHZ scheme the elements of physical reality
corresponding to these operators obey
\begin{align}\label{mxmx}
m_x^A\,\,m_x^B&= -1.\\
m_y^A\,\,m_y^B&= 1.\label{mxmx1}
\end{align}
Multiplying these two equations one obtains
\begin{align}\label{mm2}
(m_x^A\,\,m_y^A) (m_x^B\,\,m_y^B) =-1.
\end{align}
This equation agrees the local correlation (\ref{hypo1}). If fact
the local correlation is the only choice for a local theory.
Consider the following Hermitian operator
\begin{align}\label{ss}
-i\sigma_x \,\sigma_y  = \begin{pmatrix} 1 & 0 \\ 0 & -1.
\end{pmatrix}
\end{align}
It is an observable quantity with eigenvalues $\pm 1$. If elements
of physical reality do exist one has to assume
\begin{align}\label{mm3}
-im_x\, m_y =\pm 1.
\end{align}
This equation is just the local correlation (\ref{hypo1}).

\section{Local correlation in Bell's inequality}

Finally we consider Bell's inequality\cite{Aspect}, which is given
by
\begin{align}\label{bell}
f({\bf b,c})=|P({\bf a,b}) - P({\bf a, c})| + P({\bf b, c}) \leq 1,
\end{align}
where $P({\bf x,y})$ are correlation functions between the spin
components of electrons in directions ${\bf x,y=a,b,c}$. ${\bf a}$
belongs to electron A, and ${\bf b, c}$ belong to electron B.

According to quantum theory $P({\bf x,y})$ are given by
$\langle\psi|\sigma^A_{\bf x}\sigma^B_{\bf
y}|\psi\rangle=\cos(\widehat{\bf x,y})$, where $\widehat{\bf x,y}$
denote the angle between the two vectors ${\bf x,y}$. Then one has
\begin{align}
f({\bf b,c})=|\cos(\widehat{\bf a,b}) - \cos(\widehat{\bf a, c})| +
\cos(\widehat{\bf b, c}).
\end{align}

%where {\bf a, b} are two directions in which the spins of two
%entangled particles, A and B, in a spin-singlet state are measured,
%and {\bf c} is another direction in which the spin of particle B is
%measured. {\bf a, b, c} are all in a plane perpendicular to the
%velocities of the particles.
This function $f({\bf b,c})$ is plotted in FIG.2. It is seen that
(\ref{bell}) is violated  inside the four peak regions. It was
claimed that the local theory is refuted by this violation.
\begin{figure}
\includegraphics[width=7cm,height=6cm]{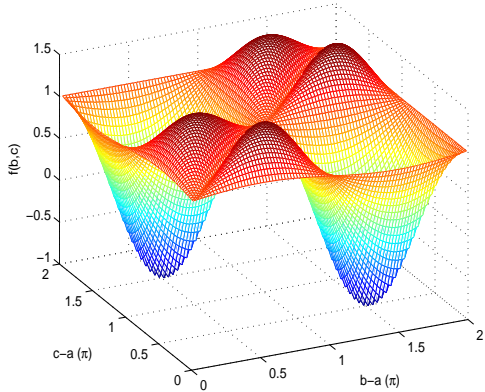}
\caption{$f({\bf b,c})$ in quantum theory. The x axis represents the
angle between {\bf b} and {\bf a}, and the y axis the angle between
{\bf c} and {\bf a}.  Violation to Bell's inequality occurs inside
the four peak regions, labeled as 1,2,3,4, where $f({\bf b,c})>1$.}
\end{figure}

In these four peak regions, however, we have $[\sigma_{\bf b},
\sigma_{\bf c}]\ne 0$, i.e., there is a local correlation between
the measured quantities in $P({\bf b,c})$. In particular, it is seen
from the figure that when $\theta\rightarrow 0$ or $\pi$, here
$\theta$ is the angle between {\bf b} and {\bf c}, the local
correlation vanishes and thus the violation disappears. A violation
without a local correlation was never found. Therefore, the
violation to Bell's inequality can be attributed to the local
correlation of one particle but not necessarily the non-locality of
the two entangled particles.

It is seen that in all the GHZ-like schemes and Bell's inequality a
common point is the local correlation between the measured
quantities on one particle. A EPR local theory can survive if  the
local correlation is  included between the elements of physical
reality. A future scheme to refute such local theories should be
built upon measurements to \textit{uncorrelated} physical
quantities.

\section{Conclusions}

In this work an EPR local theory with local correlation is proposed,
giving an explanation to the contractions in the GHZ-like schemes
for Bell's theorem and the violation to Bell's inequality. It agrees
with the predictions for the GHZ state of three  spins, two-particle
entangled state and even one spin state. It is shown that Bell's
inequality is violated only in the region of the local correlation
between the measured quantities. It is concluded that the
contradictions found in the GHZ-like schemes and the violation to
Bell's inequality can be explained by the EPR local theory with
local correlation. This local correlation, however, disobeys with
quantum theory. More stringent experiments are required to test this
EPR local theory with local correlation.

This work is supported by the Natural Science Foundation of
Guangdong province (No.7005834). Authors thank Xiaojun Wang, Xinding
Zhang and Kaihua Zhao for helpful discussions.

%\begin{acknowledgments}
%\end{acknowledgments}

%\appendix

\end{document}